\documentclass[11pt]{article}
\usepackage[centertags]{amsmath}
\usepackage{amsfonts}
\usepackage{amssymb}
\usepackage{amsthm}
\usepackage{newlfont}
\usepackage{graphicx}

\newfont{\bb}{msbm10 at 12pt}

\begin{document}

\title{Homothetic Vectors of Bianchi Type I Spacetimes in Lyra Geometry and General Relativity}
\author{ Ahmad T Ali$^{1,2}$, Suhail Khan$^{3}$ and
Azeb Alghanemi$^{1}$ \\
1- King Abdulaziz University, Faculty of Science, Department of Mathematics,\\
PO Box 80203, Jeddah, 21589, Saudi Arabia.\\
2- Mathematics Department, Faculty of Science, Al-Azhar University,\\
Nasr City, 11884, Cairo, Egypt.\\
3- Department of Mathematics, University of Peshawar, Peshawar,\\
Khyber Pakhtoonkhwa, Pakistan.\\
E-mail: suhail\_74pk@yahoo.com}


\maketitle
\begin{abstract} In this paper Bianchi type I spacetimes are completely classified by their homothetic vectors
in the context of Lyra geometry. The non-linear coupled Lyra homothetic equations are obtained and
solved completely for different cases. In some cases, Bianchi type I spacetimes admit proper Lyra homothetic vectors (LHVs) for special choices of the metric functions, while there exist other cases where the spacetime under consideration admits only Lyra Killing vectors (LKVs). In all the possible cases where Bianchi type I spacetimes admit proper LHVs or LKVs, we obtained homothetic and Killing vectors for Bianchi type I spacetimes in general relativity by taking the displacement vector of Lyra geometry as zero. Matter collineation
symmetry is explained by taking the matter field as a perfect fluid. The obtained proper LHVs and LKVs are used in matter collineation equations and a
barotropic equation of state having $\rho(t)\,=\,\gamma\,p(t)$, $0\,\leq\,\gamma\,\leq\,1$  form is always obtained when the displacement vector is considered as a function of $t$ or treated as a constant.
\end{abstract}

\emph{MSC2010:} 83F05.

\emph{Keywords}: Homothetic vectors, Lyra geometry, Matter collineations.

\section{Introduction }

Solutions of Einstein field equaitons (EFE's) are obtained through symmetries in general relativity \cite{ali1, ali2, att1, olve1,
ovsi1, yadav1}. Symmetries of a spacetime play a central role to classify these invariant solutions  \cite{khan3}.
Lie symmetries of various geometrical and physical
quantities in general relativity have been studied by many authors, e.g. see
\cite{hall1, steph1}. Different kinds of symmetries have been extensively studied and various aspects of physical and geometrical interests have been discussed (see for example \cite{gad1, hall1}).  General relativity is based upon Riemannian geometry while another equivalent description of gravity
known as teleparallel theory is based upon Weitzenbock geometry \cite{weit1} and symmetries of various kinds are also explored there \cite{khan1, khan2, shabbir1, shabbir2, shabbir3, shabbir4, shabbir5, shabbir6, shabbir7, shabbir8, sharif1,
sharif2}.
Isometries (also called Killing vectors) of a spacetimes are those
vectors which carries the metric tensor invariantly when dragged through Lie transport. Different Lie symmetries
have been used to construct new solutions of EFE's \cite{khan0}. Conformal Killing vectors always satisfy equation of the form
\begin{equation}\label{u11}
\mathcal{L}_{X}\,g_{ij}\,=\,2\,\alpha\,g_{ij},
\end{equation}
where $g_{ij}$ denotes the metric tensor components, $X$ represents
Killing vector field, $\mathcal{L}_{X}$ denotes Lie derivative
along $X$ and $\alpha$ is conformal factor depending upon spacetime coordinates
For a constant $\alpha$,
$X$ becomes a homothetic vector (HV). If this constant is \textbf{non-zero}, $X$ is called \textit{proper HV }
and for  $\alpha\,=\,0$,  $X$ becomes a \textit{Killing vector field}.
It is to remind that spacetime curvature has a great influence over CKVF's. When a spacetime remains non-conformally flat, the maximum dimension of the Lie algebra of CKVF's is seven \cite{katzin1,
katzin2}, while for conformally flat spacetimes it is fifteen.
Symmetries can be applied to explore the kinematical, dynamical or  geometrical properties of a spacetime manifold \cite{tsam1}. Conformal symmetry can be applied to different spacetime manifolds to obtain information about rotation, shear or expansion \cite{maar1}. Symmetries in general and conformal symmetry in particular have shown an important role in obtaining new solutions of EFE's \cite{cole1}.
At the geometric level symmetries make possible coordinate choice to simplify the metric \cite{tsam2}. Different modifications of Riemannian geometry are proposed to tackle the problems such as unification of the laws of electromagnetism and gravitation, coupling of matter fields with gravitational field and singularities of standard cosmology. One such modification is given by Lyra
\cite{lyra1} in 1951 by introducing a gauge function.  Riemannian geometry is also modified by Sen
\cite{sen1, sen2} and Dunn \cite{sen3}. Their proposed scalar tensor theory resembles to EFE's. Recently,  Lie derivative for tensors along vector fields on Lyra manifold is introduced in  \cite{gad2, gad3} as
\begin{equation}\label{u12}
\mathcal{L}^{L}_{X}\,g_{\mu\nu}\,=\,\dfrac{1}{x^{o}}\,\left(g_{\mu\rho}\,X^{\rho}_{\,,\nu}+g_{\rho\nu}\,X^{\rho}_{\,,\mu}\right)
+\left(g_{\mu\rho}\,\Gamma_{\lambda\nu}^{\rho}+g_{\rho\nu}\,\Gamma_{\mu\lambda}^{\rho}\right)\,X^\lambda,
\end{equation}
where $\mathcal{L}^{L}_{X}$ denotes the Lyra Lie derivative of the Lyra geometry, $x^{o}$ is a gauge function of the
spacetime coordinates and $\Gamma_{\mu\nu}^{\rho}$ is Lyra affine
connection given by:
\begin{equation}\label{u13}
\Gamma_{\mu\nu}^{\lambda}=\dfrac{1}{x^{o}}\{_{\mu\nu}^{\lambda}\}
+\dfrac{1}{2}\,\left(\delta_{\mu}^{\lambda}\,\phi_{\nu}+\delta_{\nu}^{\lambda}\,\phi_{\mu}-g_{\mu\nu}\,\phi^{\lambda}\right),
\end{equation}
where $\delta_{\mu}^{\lambda}=\mathrm{daig}(1,1,1,1)$ is the
Kronicker delta, $\{_{\mu\nu}^{\lambda}\}$ is the Riemannian
connection and $\phi_{\mu}=\left(\theta(t),0,0,0\right)$ is
displacement vector with a displacement function $\theta(t)$.

Homothetic symmetries in Lyra geometry have been discussed for only few spactimes by Gad and Alofi
\cite{gad2} and Gad \cite{gad3}. They obtained only special solutions
of their homothetic equations. Keeping in mind the wide range applications of Lie symmetries at kinematics, dynamics and  geometric levels  in
general theory of relativity, our aim is to completely classify Bianchi type I spacetimes by its  LHVs. Also, proper LHVs and LKVs will be used into matter collineations equations to obtain a barotropic equation of state of the form $\rho(t)\,=\,\gamma\,p(t)$, $0\,\leq\,\gamma\,\leq\,1$ for the perfect fluid matter field.\\
It is important to highlight that this paper supersedes \cite {ali3} by improving some results and by introducing a new section of Matter Collineation. Some notations are also changed in this new version. The remainder part of this paper is organized as follows: In section 2, non-linear
coupled Lyra homothetic equations  are
introduced. Those equations are integrated directly to obtain a general solution. During the process some integrability conditions are
also obtained, which are also solved. In section 3, matter collineations equations are obtained and solved to
get a barotropic equation of state. In the last section a conclusion to the paper is given.

\section{Lyra homothetic equations and their solutions}
We consider Bianchi type I spacetimes in the convention
coordinates $(x^0=t,\,x^1=x,\,x^2=y,\,x^3=z)$, in the form
\begin{equation}  \label{u31}
ds^2\,=\,-dt^2+U^2(t)\,dx^2+W^2(t)\,\left(dy^2+dz^2\right),
\end{equation}
where $U$ and $W$ are no where zero functions of $t$ only. If we choose the normal
gauge $x^{o}=1$, the non-zero Lyra connections for spacetime (\ref{u31}) can be computed through Eq.(\ref{u13}) as follows
\begin{equation}\label{u32}
\left\{
\begin{array}{ll}
\Gamma_{00}^{0}\,=\,\dfrac{\theta(t)}{2},\,\,\,\,\,\,\,\,\,\,\,\,\,\,\,\,\,
\Gamma_{01}^1\,=\,\dfrac{U'(t)}{U(t)}+\dfrac{\theta(t)}{2},\,\,\,\,\,\,\,\,\,\,\,\,\,\,\,\,\,
\Gamma_{02}^2\,=\,\Gamma_{03}^3\,=\,\dfrac{W'(t)}{W(t)}+\dfrac{\theta(t)}{2},\\
\\
\Gamma_{11}^0\,=\,U(t)\,U'(t)+\dfrac{U^2(t)\,\theta(t)}{2},\,\,\,\,\,\,\,\,\,\,\,\,\,\,\,\,
\Gamma_{22}^0\,=\,\Gamma_{33}^0\,=\,W(t)\,W'(t)+\dfrac{W^2(t)\,\theta(t)}{2},
\end{array}
\right.
\end{equation}
where prime denotes the derivatives with respect to $t$. Using the
metric (\ref{u31}) and Lyra connections (\ref{u32}) in the Lyra
homothetic equation (\ref{u11}) along with Lyra Lie derivative
(\ref{u12}) yields the system of ten partial differential equations
in four unknowns $X^0$, $X^1$, $X^2$ and $X^3$ as the following:
\begin{equation}  \label{u35-1}
2\,X^0_{\,,t}+\theta(t)\,X^0\,=\,2\,\alpha,
\end{equation}
\begin{equation}  \label{u35-2}
X^0_{\,,x}\,=\,U^2\,X^1_{\,,t},
\end{equation}
\begin{equation}  \label{u35-3}
X^0_{\,,y}\,=\,W^2\,X^2_{\,,t},
\end{equation}
\begin{equation}  \label{u35-4}
X^0_{\,,z}\,=\,W^2\,X^3_{\,,t},
\end{equation}
\begin{equation}  \label{u35-5}
2\,U(t)\,X^1_{\,,x}+\left[\theta(t)\,U(t)+2\,U'(t)\right]\,X^0\,=\,2\,\alpha\,U(t),
\end{equation}
\begin{equation}  \label{u35-6}
X^{1}_{\,,y}\,U^2+W^2\,X^2_{\,,x}\,=\,0,
\end{equation}
\begin{equation}  \label{u35-7}
X^{1}_{\,,z}\,U^2+W^2\,X^3_{\,,x}\,=\,0,
\end{equation}
\begin{equation}  \label{u35-8}
2\,W(t)\,X^2_{\,,y}+\left[\theta(t)\,W(t)+2\,W'(t)\right]\,X^0\,=\,2\,\alpha\,W(t),
\end{equation}
\begin{equation}  \label{u35-9}
X^{2}_{\,,z}+X^3_{\,,y}\,=\,0,
\end{equation}
\begin{equation}  \label{u35-10}
2\,W(t)\,X^3_{\,,z}+\left[\theta(t)\,W(t)+2\,W'(t)\right]\,X^0\,=\,2\,\alpha\,W(t),
\end{equation}
We will integrate some suitable equations from the above equations to get vector field components
$X^0$, $X^1$, $X^2$ and $X^3$. The whole
process is explained as follows:
Differentiating Eq. (\ref{u35-3}) with respect to $z$, Eq.
(\ref{u35-4}) with respect to $y$ and Eq. (\ref{u35-9}) with respect
to $t$ to obtain a relation
\begin{equation}  \label{u41-1}
X^0_{\,,yz}\,=\,X^2_{\,,tz}\,=\,X^3_{\,,ty}\,=\,0.
\end{equation}
Similarly, differentiating Eqs. (\ref{u35-6}), (\ref{u35-7}) and (\ref{u35-9}),
 with respect to $z$, $y$, $x$, respectively and simplifying, we get
\begin{equation}  \label{u41-2}
X^1_{\,,yz}\,=\,X^2_{\,,xz}\,=\,X^3_{\,,xy}\,=\,0.
\end{equation}
By applying similar techniques between Eqs. (\ref{u35-3}),
(\ref{u35-4}), (\ref{u35-6}), (\ref{u35-7}), (\ref{u35-8}),
(\ref{u35-9}) and (\ref{u35-10}), one gets
\begin{equation}  \label{u41-3}
X^i_{\,,yy}\,-\,X^i_{\,,zz}\,=\,X^j_{\,,yy}\,+\,X^j_{\,,zz}\,=\,0,\,\,\,i=0,1,\,\,\,j=2,3.
\end{equation}
Integrating Eqs. (\ref{u41-1}) and (\ref{u41-2}), substituting the
results in (\ref{u41-3}), integrating again (\ref{u41-3}) and solving with the help of Eqs. (\ref{u35-3}), (\ref{u35-4}), (\ref{u35-6}), (\ref{u35-7})
and (\ref{u35-9}), we get
\begin{equation}  \label{u41-4}
\left\{
  \begin{array}{ll}
    X^0\,=\,F^0+W^2(t)\,\Big[y\,F^2+z\,F^3+\left(y^2+z^2\right)\,F^4\Big]_{,t},\\
\\
  X^1\,=\,\alpha\,x+F^1-\dfrac{W^2(t)}{U^2(t)}\,\Big[y\,F^2+z\,F^3+\left(y^2+z^2\right)\,F^4\Big]_{,x},\\
\\
  X^2\,=\,F^2+(\alpha+2\,F^4)\,y+d_1\,z+d_2\,\left(y^2-z^2\right)+2\,d_3\,y\,z,\\
\\
X^3\,=\,F^3-d_1\,y+(\alpha+2\,F^4)\,z-d_3\,\left(y^2-z^2\right)+2\,d_2\,y\,z,
  \end{array}
\right.
\end{equation}
$F^i=F^i(t,x)$ are integration functions for all
$i=0,1,2,3,4$, while $d_1$, $d_2$ and $d_3$ are constants of
integration. Substituting the above vector field components in the system (\ref{u35-1})-(\ref{u35-10}) and making use of  \textit{Mathematica Program} we get the
following constraints:
\begin{equation}  \label{u410-1}
W(t)\,\left[\theta(t)\,W(t)+2\,W'(t)\right]F^{i}_{\,,t}+4\,d_i\,=\,0,\,\,\,\,\,\,\,\,\,\,i=2,3,4,
\end{equation}
\begin{equation}  \label{u410-2}
\left[\theta(t)\,W(t)+2\,W'(t)\right]F^{0}+4\,W(t)\,F^4\,=\,0,
\end{equation}
\begin{equation}  \label{u410-3}
U(t)\,\left[\theta(t)\,U(t)+2\,U'(t)\right]F^{i}_{\,,t}\,=\,2\,F^{i}_{\,,xx},\,\,\,\,\,\,\,\,\,\,i=2,3,4,
\end{equation}
\begin{equation}  \label{u410-4}
\left[\theta(t)\,U(t)+2\,U'(t)\right]F^{0}+2\,U(t)\,F^1_{\,,x}\,=\,0,
\end{equation}
\begin{equation}  \label{u410-5}
\left(\left[\dfrac{W(t)}{U(t)}\right]\,\left[\dfrac{U(t)}{W(t)}\right]'\,F^i-F^i_{\,,t}\right)_{\,,x}\,=\,0,\,\,\,\,\,i=2,3,4,
\end{equation}
\begin{equation}  \label{u410-6}
F^0_{\,,x}\,=\,U^2(t)\,F^1_{\,,t},
\end{equation}
\begin{equation}  \label{u410-7}
\left[\theta(t)\,W(t)+4\,W'(t)\right]F^{i}_{\,,t}+2\,W(t)\,F^{i}_{\,,tt}\,=\,0,\,\,\,\,\,\,\,\,\,\,i=2,3,4,
\end{equation}
\begin{equation}  \label{u410-8}
\theta(t)\,F^0+2\,F^0_{\,,t}\,=\,2\,\alpha,
\end{equation}
where $d_4\,=\,0$. We have solved these constrains for different choices of the metric functions and  displacement
function $\theta(t)$ and the final form
of LHV's are obtained. In the following, only final results are listed to avoid lengthy details:

\subsection{Proper Lyra Homothetic Vectors For Bianchi Type I Spacetimes }
In this section different possibilities for the metric functions and
displacement function are explored where Bianchi type I spacetimes admit proper Lyra homothetic vectors. The cases
where the spacetimes manifolds admit only Lyra Killing vectors will be discussed in the next sub-section. Details are omitted
and the results are written here directly as different possibilities:\\

\textbf{Solution (LG1):}
\begin{equation}  \label{u41-HS-13}
\left\{
  \begin{array}{ll}
    X^0\,=\,\exp\Big[-\int\dfrac{\theta(t)}{2}\,dt\Big]\,\left(a_1+a_2\,x+a_3\,y+a_4\,z+\alpha\,\int\exp\Big[\int\dfrac{\theta(t)}{2}\,dt\Big]\,dt\right),\\
\\
X^1\,=\,a_6+\alpha\,x+a_7\,y+a_8\,z+a_2\,\int\exp\Big[\int\dfrac{\theta(t)}{2}\,dt\Big]\,dt,\\
\\
X^2\,=\,a_9-a_7\,x+\alpha\,y+a_{10}\,z+a_3\,\int\exp\Big[\int\dfrac{\theta(t)}{2}\,dt\Big]\,dt,\\
\\
X^3\,=\,a_{11}-a_8\,x-a_{10}\,y+\alpha\,z+a_4\,\int\exp\Big[\int\dfrac{\theta(t)}{2}\,dt\Big]\,dt,
  \end{array}
\right.
\end{equation}
where $U(t)\,=\,W(t)\,=\,\exp\Big[-\int\dfrac{\theta(t)}{2}\,dt\Big]$ while $\theta(t)$ is an arbitrary function, $a_i$, $i\,=\,1,2,...,11$, are arbitrary constants such that $a_5\,=\,\alpha$.\\

Subtracting Killing vector fields from (\ref{u41-HS-13}), the proper Lyra homothetic vector is obtained as
\begin{equation}  \label{u41-HS-13-P}
X\,=\,\exp\Big[-\int\dfrac{\theta(t)}{2}\,dt\Big]\,\left(\int\exp\Big[\int\dfrac{\theta(t)}{2}\,dt\Big]\,dt\right)\,\dfrac{\partial}{\partial t}+x\,\dfrac{\partial}{\partial x}+y\,\dfrac{\partial}{\partial y}+z\,\dfrac{\partial}{\partial z}.
\end{equation}

The generators of Lyra homothetic vector fields (\ref{u41-HS-13}) can be written as the following:
\begin{equation}  \label{u41-HS-13-1}
X\,=\,\sum_{i=1}^{11}\,a_i\,Z_i,
\end{equation}
where
\begin{equation}  \label{u41-HS-13-2}
\left\{
  \begin{array}{ll}
    Z_1\,=\,\exp\Big[-\int\dfrac{\theta(t)}{2}\,dt\Big]\,\,\dfrac{\partial}{\partial t},\\
    \\
    Z_2\,=\,x\,\exp\Big[-\int\dfrac{\theta(t)}{2}\,dt\Big]\,\dfrac{\partial}{\partial t}+\left(\int\exp\Big[\int\dfrac{\theta(t)}{2}\,dt\Big]\,dt\right)\,\dfrac{\partial}{\partial x},\\
    \\
    Z_3\,=\,y\,\exp\Big[-\int\dfrac{\theta(t)}{2}\,dt\Big]\,\,\dfrac{\partial}{\partial t}+\left(\int\exp\Big[\int\dfrac{\theta(t)}{2}\,dt\Big]\,dt\right)\,\dfrac{\partial}{\partial y},\\
    \\
    Z_4\,=\,z\,\exp\Big[-\int\dfrac{\theta(t)}{2}\,dt\Big]\,\,\dfrac{\partial}{\partial t}+\left(\int\exp\Big[\int\dfrac{\theta(t)}{2}\,dt\Big]\,dt\right)\,\dfrac{\partial}{\partial z},\\
    \\
    Z_5\,=\,\exp\Big[-\int\dfrac{\theta(t)}{2}\,dt\Big]\,\left(\int\exp\Big[\int\dfrac{\theta(t)}{2}\,dt\Big]\,dt\right)\,\dfrac{\partial}{\partial t}+x\,\dfrac{\partial}{\partial x}+y\,\dfrac{\partial}{\partial y}+z\,\dfrac{\partial}{\partial z},\\
    \\
    Z_6\,=\,\dfrac{\partial}{\partial x},\,\,\,\,\,\,\,\,\,\,\,\,\,\,\,
    Z_7\,=\,y\,\dfrac{\partial}{\partial x}-x\,\dfrac{\partial}{\partial y},\,\,\,\,\,\,\,\,\,\,\,\,\,\,\,
        Z_8\,=\,z\,\dfrac{\partial}{\partial x}-x\,\dfrac{\partial}{\partial z},\\
    \\
    Z_9\,=\,\dfrac{\partial}{\partial y},\,\,\,\,\,\,\,\,\,\,\,\,\,\,\,
    Z_{10}\,=\,z\,\dfrac{\partial}{\partial y}-y\,\dfrac{\partial}{\partial z},\,\,\,\,\,\,\,\,\,\,\,\,\,\,\,
    z_{11}\,=\,\dfrac{\partial}{\partial z}.
  \end{array}
\right.
\end{equation}
The generators above form close Lie algebra structure with non-zero Lie brackets, given by:
\begin{equation}  \label{u41-HS-13-3}
\left\{
  \begin{array}{ll}
    \left[Z_1,Z_2\right]\,=\,Z_6,\,\,\,\,\,\,\,\left[Z_1,Z_3\right]\,=\,Z_9,\,\,\,\,\,\,\,\left[Z_1,Z_4\right]\,=\,Z_{11},\,\,\,\,\,\,\,\left[Z_1,Z_5\right]\,=\,Z_{1},\\
    \\
    \left[Z_2,Z_3\right]\,=\,-Z_7,\,\,\,\,\,\,\,\left[Z_2,Z_4\right]\,=\,-Z_8,\,\,\,\,\,\,\,\left[Z_2,Z_6\right]\,=\,Z_{1},\,\,\,\,\,\,\,\left[Z_2,Z_7\right]\,=\,-Z_{3},\\
    \\
    \left[Z_2,Z_8\right]\,=\,-Z_{4},\,\,\,\,\,\,\,\left[Z_3,Z_4\right]\,=\,-Z_{10},\,\,\,\,\,\,\,\left[Z_3,Z_7\right]\,=\,Z_{2},\,\,\,\,\,\,\,\left[Z_3,Z_9\right]\,=\,-Z_{1},\\
    \\
    \left[Z_3,Z_{10}\right]\,=\,-Z_{4},\,\,\,\,\,\,\,\left[Z_4,Z_8\right]\,=\,Z_{2},\,\,\,\,\,\,\,\left[Z_4,Z_{10}\right]\,=\,Z_{3},\,\,\,\,\,\,\,\left[Z_4,Z_{11}\right]\,=\,-Z_{1},\\
    \\
    \left[Z_5,Z_{6}\right]\,=\,-Z_{6},\,\,\,\,\left[Z_5,Z_9\right]\,=\,-Z_{9},\,\,\,\,\left[Z_5,Z_{11}\right]\,=\,-Z_{11},\,\,\,\,\left[Z_6,Z_{7}\right]\,=\,-Z_{9},\\
    \\
    \left[Z_6,Z_{8}\right]\,=\,-Z_{11},\,\,\,\,\left[Z_7,Z_8\right]\,=\,Z_{10},\,\,\,\,\left[Z_7,Z_{9}\right]\,=\,-Z_{6},\,\,\,\,\left[Z_7,Z_{10}\right]\,=\,-Z_{8},\\
    \\
    \left[Z_8,Z_{10}\right]\,=\,Z_{7},\,\,\,\,\left[Z_8,Z_{11}\right]\,=\,-Z_{6},\,\,\,\,\left[Z_9,Z_{10}\right]\,=\,-Z_{11},\,\,\,\,\left[Z_{10},Z_{11}\right]\,=\,-Z_{9}
  \end{array}
\right.
\end{equation}

Later, in Section 3, the role of the displacement vector $\theta$ will be analyzed in determining the Matter Collineation symmetry
and in the investigation of barotropic equation of state. In a paper \cite{gad3}, the author claimed that this equation of
state never satisfies when $\theta$ remians a function of $t$ or a constant. It is claimed that it satisfies only when $\theta=0$. In order to check whether or not barotropic equation of state satisfies in our case, we need to
obtain homothetic vectors when $\theta$ is a constant and when  $\theta=0$. These two solutions are obtained in the following:

\textbf{Solution (GR1): ($\theta\,=\,0$):}
It is interesting to see that the metric functions are dependent upon the displacement
funciton $\theta$ and taking $\theta=0$, the Bianchi type I spacetime becomes flat and its LHVs reduce to the
HVs of general relativity \cite{shabbir9}, which are given as follows:

\begin{equation}  \label{u41-HS-13-GR}
\left\{
  \begin{array}{ll}
    X^0\,=\,a_1+\alpha\,t+a_2\,x+a_3\,y+a_4\,z,\\
\\
X^1\,=\,a_6+a_2\,t+\alpha\,x+a_7\,y+a_8\,z,\\
\\
X^2\,=\,a_9+a_3\,t-a_7\,x+\alpha\,y+a_{10}\,z,\\
\\
X^3\,=\,a_{11}+a_4\,t-a_8\,x-a_{10}\,y+\alpha\,z,
  \end{array}
\right.
\end{equation}
where $U(t)\,=\,W(t)\,=\,1$ and $a_i$, $i\,=1,2,...,11$ are constants such that $a_5\,=\,\alpha$.
The proper homothetic vector field is obtained as:
\begin{equation}  \label{u41-HS-13-P-GR}
X\,=\,t\,\dfrac{\partial}{\partial t}+x\,\dfrac{\partial}{\partial x}+y\,\dfrac{\partial}{\partial y}+z\,\dfrac{\partial}{\partial z}.
\end{equation}

\textbf{Solution (LG1): ($\theta\,=\,2\,\theta_0$):}
This solution is obtained from the above case $(LG1)$ by taking $\theta$ as constant.
\begin{equation}  \label{u41-HS-13-C}
\left\{
  \begin{array}{ll}
    X^0\,=\,\left(a_1+a_2\,x+a_3\,y+a_4\,z\right)\,e^{-\theta_0\,t}+\dfrac{\alpha}{\theta_0},\\
\\
X^1\,=\,a_6+\alpha\,x+a_7\,y+a_8\,z+a_2\,e^{\theta_0\,t},\\
\\
X^2\,=\,a_9-a_7\,x+\alpha\,y+a_{10}\,z+a_3\,e^{\theta_0\,t},\\
\\
X^3\,=\,a_{11}-a_8\,x-a_{10}\,y+\alpha\,z+a_4\,e^{\theta_0\,t},
  \end{array}
\right.
\end{equation}
where $U(t)\,=\,W(t)\,=\,e^{-\theta_0\,t}$, $a_i$, $i\,=\,1,2,...,11$, are constants obtained during integration such that $a_5\,=\,\alpha$.
In this case the proper Lyra homothetic vector becomes:
\begin{equation}  \label{u41-HS-13-P-C}
X\,=\,\dfrac{1}{\theta_0}\,\dfrac{\partial}{\partial t}+x\,\dfrac{\partial}{\partial x}+y\,\dfrac{\partial}{\partial y}+z\,\dfrac{\partial}{\partial z}.
\end{equation}

\textbf{Solution (LG2):}
\begin{equation}  \label{0u41-HS-13}
\left\{
  \begin{array}{ll}
    X^0\,=\,\exp\Big[-\int\dfrac{\theta(t)}{2}\,dt\Big]\,\left(a_1+\alpha\,\int\exp\Big[\int\dfrac{\theta(t)}{2}\,dt\Big]\,dt\right),\\
\\
X^1\,=\,a_2+a_3\,x+a_4\,y+a_5\,z,\\
\\
X^2\,=\,a_6-a_4\,x+a_3\,y+a_5\,z,\\
\\
X^3\,=\,a_8-a_5\,x-a_7\,y+a_3\,z,
  \end{array}
\right.
\end{equation}
where $U(t)\,=\,W(t)\,=\,\exp\Big[-\int\dfrac{\theta(t)}{2}\,dt\Big]\,\left(a_1+\alpha\,\int\exp\Big[\int\dfrac{\theta(t)}{2}\,dt\Big]\,dt\right)^{1-
    \dfrac{a_3}{\alpha}}$ while $\theta(t)$ is an arbitrary function, $\alpha$ and $a_i$, $i\,=\,1,2,...,8$, are arbitrary constants such that $a_3\,\neq\,\alpha$.\\

Subtracting Killing vector fields from (\ref{0u41-HS-13}), the proper Lyra homothetic vector is obtained as
\begin{equation}  \label{u41-HS-13-P}
X\,=\,\exp\Big[-\int\dfrac{\theta(t)}{2}\,dt\Big]\,\left(\int\exp\Big[\int\dfrac{\theta(t)}{2}\,dt\Big]\,dt\right)\,\dfrac{\partial}{\partial t}.
\end{equation}

\textbf{Solution (GR2): ($\theta\,=\,0$):}
It is interesting to see that the metric functions are dependent upon the displacement
funciton $\theta$ and taking $\theta=0$, the LHVs reduce to the
HVs of general relativity, as obtained in \cite{shabbir9}, case (5) and are given as follows:
\begin{equation}  \label{0u41-HS-13-GR}
\left\{
  \begin{array}{ll}
    X^0\,=\,a_1+\alpha\,t,\\
\\
X^1\,=\,a_2+a_3\,x+a_4\,y+a_5\,z,\\
\\
X^2\,=\,a_6-a_4\,x+a_3\,y+a_5\,z,\\
\\
X^3\,=\,a_8-a_5\,x-a_7\,y+a_3\,z,
  \end{array}
\right.
\end{equation}
where $U(t)\,=\,W(t)\,=\,\left(a_1+\alpha\,t\right)^{1-
    \dfrac{a_3}{\alpha}}$ and $a_i$, $i\,=\,1,2,...,8$ are constants such that $a_3\,\neq\,\alpha$.
The proper homothetic vector field is obtained as:
\begin{equation}  \label{0u41-HS-13-P-GR}
X\,=\,t\,\dfrac{\partial}{\partial t}.
\end{equation}

\textbf{Solution (LG2): ($\theta\,=\,2\,\theta_0$):}
This solution is obtained from the above case $(LG2)$ by taking $\theta$ as constant.
\begin{equation}  \label{0u41-HS-13-C}
\left\{
  \begin{array}{ll}
    X^0\,=\,a_1\,e^{-\theta_0\,t}+\dfrac{\alpha}{\theta_0},\\
\\
X^1\,=\,a_2+a_3\,x+a_4\,y+a_5\,z,\\
\\
X^2\,=\,a_6-a_4\,x+a_3\,y+a_5\,z,\\
\\
X^3\,=\,a_8-a_5\,x-a_7\,y+a_3\,z,
  \end{array}
\right.
\end{equation}
where $U(t)\,=\,W(t)\,=\,e^{-\theta_0\,t}\,\left(a_1+\dfrac{\alpha}{\theta_0}\,e^{\theta_0\,t}\right)^{1-\dfrac{a_3}{\alpha}}$, $a_i$, $i\,=\,1,2,...,8$, are constants obtained during integration such that $a_3\,\neq\,\alpha$.
In this case the proper Lyra homothetic vector becomes:
\begin{equation}  \label{u41-HS-13-P-C}
X\,=\,\dfrac{1}{\theta_0}\,\dfrac{\partial}{\partial t}.
\end{equation}

\textbf{Solution (LG3):}
\begin{equation}  \label{u41-HS-8}
\left\{
  \begin{array}{ll}
    X^0\,=\,W(t)\Bigg[\left(a_1+a_2\,y+a_3\,z\right)\,e^{a_0\,x}+\left(a_4+a_5\,y+a_6\,z\right)\,e^{-a_0\,x}\\
    \,\,\,\,\,\,\,\,\,\,\,\,\,\,\,\,\,\,\,\,\,\,\,\,\,\,\,\,\,\, \,\,\,\,\,\,\,\,\,\,\,\,\,\,\,\,\,\,\,\,
    \,\,\,\,\,\,\,\,\,\,\,\,\,\,\,\,\,\,\,\,\,\,\,\,\,\,\,\,\,\,\,\,\,\,\,\,\,\,\,\,
    +\alpha\,\left(\gamma_0+\int\exp\Big[\int\dfrac{\theta(t)}{2}\,dt\Big]\,dt\right)\Bigg],\\
\\
X^1=a_7-a_0^{-1}\,\Big[\left(a_1+a_2y+a_3z\right)\mathrm{e}^{a_0\,x}-\left(a_4+a_5y+a_6z\right)\mathrm{e}^{-a_0\,x}\Big]\,\\
\,\,\,\,\,\,\,\,\,\,\,\,\,\,\,\,\,\,\,\,\,\,\,\,\,\,\,\,\,\, \,\,\,\,\,\,\,\,\,\,\,\,\,\,\,\,\,\,\,\,
\,\,\,\,\,\,\,\,\,\,\,\,\,\,\,\,\,\,\,\,\,\,\,\,\,\,\,\,\,\,\,\,\,\,\,\,\,\,\,\,
\times\,\left(\gamma_0+\int\exp\Big[\int\dfrac{\theta(t)}{2}\,dt\Big]\,dt\right)^{-1},\\
\\
X^2\,=\,a_8+\alpha\,y+a_9\,z+\Big(a_2\,\,e^{a_0\,x}+a_5\,e^{-a_0\,x}\Big)\,
\left(\gamma_0+\int\exp\Big[\int\dfrac{\theta(t)}{2}\,dt\Big]\,dt\right),\\
\\
X^3\,=\,a_{10}-a_9\,y+\alpha\,z+\Big(a_3\,\,e^{a_0\,x}+a_6\,e^{-a_0\,x}\Big)\,
\left(\gamma_0+\int\exp\Big[\int\dfrac{\theta(t)}{2}\,dt\Big]\,dt\right),
  \end{array}
\right.
\end{equation}
where $U(t)\,=\,a_0\,W(t)\,\left(\gamma_0+\int\exp\Big[\int\dfrac{\theta(t)}{2}\,dt\Big]\,dt\right)$
and $W(t)\,=\,\exp\Big[-\int\dfrac{\theta(t)}{2}\,dt\Big]$ while $\theta(t)$ is an
arbitrary function and $\alpha$, $\gamma_0$, $a_i$, $i=0,1,...,10$
are arbitrary constants such that $a_0\,\neq\,0$.

Subtracting Killing vector fields from (\ref{u41-HS-8}), the proper Lyra homothetic vector field is obtained as
\begin{equation}  \label{u41-HS-8-P}
X\,=\,\exp\Big[-\int\dfrac{\theta(t)}{2}\,dt\Big]\,\left(\gamma_0+\int\exp\Big[\int\dfrac{\theta(t)}{2}\,dt\Big]\,dt\right)\,\dfrac{\partial}{\partial t}+y\,\dfrac{\partial}{\partial y}+z\,\dfrac{\partial}{\partial z}.
\end{equation}

\textbf{Solution (GR3): ($\theta\,=\,0$):} For the above case LG3, the HVFs of general relativity can be obtained by taking $\theta\,=\,0$ as follows:
\begin{equation}  \label{u41-HS-8-GR}
\left\{
  \begin{array}{ll}
    X^0\,=\,\left(a_1+a_2\,y+a_3\,z\right)\,e^{x}+\left(a_4+a_5\,y+a_6\,z\right)\,e^{-x}+\alpha\,\left(\gamma_0+t\right),\\
\\
X^1=a_7+\Big[\left(a_4+a_5y+a_6z\right)\mathrm{e}^{-x}-\left(a_1+a_2y+a_3z\right)\mathrm{e}^{x}\Big]\,\left(\gamma_0+t\right)^{-1},\\
\\
X^2\,=\,a_8+\alpha\,y+a_9\,z+\Big(a_2\,\,e^{x}+a_5\,e^{-x}\Big)\,
\left(\gamma_0+t\right),\\
\\
X^3\,=\,a_{10}-a_9\,y+\alpha\,z+\Big(a_3\,\,e^{x}+a_6\,e^{-x}\Big)\,
\left(\gamma_0+t\right),
  \end{array}
\right.
\end{equation}
where $a_0\,=\,1$, $U(t)\,=\,\gamma_0+t$, $W(t)\,=\,1$ and $\theta(t)\,=\,0$ while $\alpha$, $\gamma_0$, $a_i$, $i=1,...,10$
are arbitrary constants. To the best of our knowledge, this result is not obtained previously in literature. Subtracting Killing vector fields from (\ref{u41-HS-8-GR}), the proper homothetic vector of general relativity can be obtained as
\begin{equation}  \label{u41-HS-8-P-GR}
X\,=\,\left(\gamma_0+t\right)\,\dfrac{\partial}{\partial t}+y\,\dfrac{\partial}{\partial y}+z\,\dfrac{\partial}{\partial z}.
\end{equation}
When the displacement function become constant, that is when $\theta(t)$ is
an arbitrary constant equals $2\,\theta_0$,  homothetic vector fields in Lyra geometry
corresponding to solution \textbf{(LG3)} takes the form:\\
\textbf{Solution (LG3): ($\theta\,=\,2\,\theta_0$):}
\begin{equation}  \label{u41-HS-8-C}
\left\{
  \begin{array}{ll}
    X^0\,=\,e^{-\theta_0\,t}\Big[\left(a_1+a_2\,y+a_3\,z\right)\,e^{x}+\left(a_4+a_5\,y+a_6\,z\right)\,e^{-x}\Big]
    +\alpha\,\left(\gamma_0\,e^{-\theta_0\,t}+\theta_0^{-1}\right),\\
\\
X^1=a_7+\Big[\left(a_4+a_5y+a_6z\right)\mathrm{e}^{-x}-\left(a_1+a_2y+a_3z\right)\mathrm{e}^{x}\Big]\,
\left(\gamma_0+\theta_0^{-1}\,e^{\theta_0\,t}\right)^{-1},\\
\\
X^2\,=\,a_8+\alpha\,y+a_9\,z+\Big(a_2\,\,e^{x}+a_5\,e^{-x}\Big)\,
\left(\gamma_0+\theta_0^{-1}\,e^{\theta_0\,t}\right),\\
\\
X^3\,=\,a_{10}-a_9\,y+\alpha\,z+\Big(a_3\,\,e^{x}+a_6\,e^{-x}\Big)\,
\left(\gamma_0+\theta_0^{-1}\,e^{\theta_0\,t}\right),
  \end{array}
\right.
\end{equation}
where $a_0\,=\,1$, $U(t)\,=\,\gamma_0\,e^{-\theta_0\,t}+\theta_0^{-1}$
and $W(t)\,=\,e^{-\theta_0\,t}$ while $\alpha$, $\gamma_0$, $\theta_0$, $a_i$, $i=1,...,10$
are arbitrary constants.

Subtracting Killing vector fields from (\ref{u41-HS-8-C}), the proper Lyra homothetic vector is obtained as
\begin{equation}  \label{u41-HS-13-P-C}
X\,=\,\left(\gamma_0\,e^{-\theta_0\,t}+\theta_0^{-1}\right)\,\dfrac{\partial}{\partial t}+x\,\dfrac{\partial}{\partial x}+y\,\dfrac{\partial}{\partial y}+z\,\dfrac{\partial}{\partial z}.
\end{equation}

\textbf{Solution (LG4):}
\begin{equation}  \label{u41-HS-12}
\left\{
  \begin{array}{ll}
    X^0\,=\,\left(a_1+\alpha\,\int\exp\Big[\dfrac{1}{2}\,\int\theta(t)\,dt\Big]\,dt\right)\,\exp\Big[-\dfrac{1}{2}\,\int\theta(t)\,dt\Big],\\
    \\
X^1\,=\,a_3+a_2\,x,
\,\,\,\,\,\,\,\,\,\,\,\,\,\,\,\,
X^2=a_4+a_5\,y+a_6\,z,
\,\,\,\,\,\,\,\,\,\,\,\,\,\,\,\,
X^3=a_7-a_6\,y+a_5\,z,
  \end{array}
\right.
\end{equation}
where $\alpha$, $a_i$, $i=1,...,7$ are arbitrary
constants such that $\alpha\,\neq\,0$ while $\theta(t)$ is an arbitrary
function. For different possibilities, the proper Lyra homothetic vectors can be obtained as the following along with the metric functions:\\

\textbf{(1):} If $a_2\,\neq\,\alpha$ and $a_5\,\neq\,\alpha$, then
\begin{equation}  \label{u41-HS-12-P}
X\,=\,\exp\Big[-\dfrac{1}{2}\,\int\theta(t)\,dt\Big]\,\left(\int\exp\Big[\dfrac{1}{2}\,\int\theta(t)\,dt\Big]\,dt\right)\,\dfrac{\partial}{\partial t},
\end{equation}
where $$U(t)=a_0\,\exp\Big[-\dfrac{1}{2}\,\int\theta(t)\,dt\Big]\,\left(a_1+\alpha\,\int\exp\Big[\dfrac{1}{2}\,\int\theta(t)\,dt\Big]\,dt\right)^{1-a_2/\alpha}$$
and $$W(t)=b_0\exp\Big[-\dfrac{1}{2}\,\int\theta(t)\,dt\Big]\,\left(a_1+\alpha\,\int\exp\Big[\dfrac{1}{2}\,\int\theta(t)\,dt\Big]\,dt\right)^{1-a_5/\alpha}$$ while $a_0$ and $b_0$ are arbitrary constants.\\

\textbf{(2):} If $a_2\,=\,\alpha$ and $a_5\,\neq\,\alpha$, then
\begin{equation}  \label{u41-HS-12-P}
X\,=\,\exp\Big[-\dfrac{1}{2}\,\int\theta(t)\,dt\Big]\,\left(\int\exp\Big[\dfrac{1}{2}\,\int\theta(t)\,dt\Big]\,dt\right)\,\dfrac{\partial}{\partial t}+x\,\dfrac{\partial}{\partial x}.
\end{equation}
where $$W(t)=b_0\exp\Big[-\dfrac{1}{2}\,\int\theta(t)\,dt\Big]\,\left(a_1+\alpha\,\int\exp\Big[\dfrac{1}{2}\,\int\theta(t)\,dt\Big]\,dt\right)^{1-a_5/\alpha}$$ and $U(t)=a_0\,\exp\Big[-\dfrac{1}{2}\,\int\theta(t)\,dt\Big]$ while $a_0$ and $b_0$ are arbitrary constants.\\

\textbf{(3):} If $a_2\,\neq\,\alpha$ and $a_5\,=\,\alpha$, then
\begin{equation}  \label{u41-HS-12-P}
X\,=\,\exp\Big[-\dfrac{1}{2}\,\int\theta(t)\,dt\Big]\,\left(\int\exp\Big[\dfrac{1}{2}\,\int\theta(t)\,dt\Big]\,dt\right)\,\dfrac{\partial}{\partial t}
+y\,\dfrac{\partial}{\partial y}+z\,\dfrac{\partial}{\partial z}.\\
\end{equation}
where $$U(t)=a_0\,\exp\Big[-\dfrac{1}{2}\,\int\theta(t)\,dt\Big]\,\left(a_1+\alpha\,\int\exp\Big[\dfrac{1}{2}\,\int\theta(t)\,dt\Big]\,dt\right)^{1-a_2/\alpha}$$
and $W(t)=b_0\exp\Big[-\dfrac{1}{2}\,\int\theta(t)\,dt\Big]$ while $a_0$ and $b_0$ are arbitrary constants.

\textbf{(4):} If $a_2\,=\,\alpha$ and $a_5\,=\,\alpha$, then
\begin{equation}  \label{u41-HS-12-P}
X\,=\,\exp\Big[-\dfrac{1}{2}\,\int\theta(t)\,dt\Big]\,\left(\int\exp\Big[\dfrac{1}{2}\,\int\theta(t)\,dt\Big]\,dt\right)\,\dfrac{\partial}{\partial t}
+x\,\dfrac{\partial}{\partial x}
+y\,\dfrac{\partial}{\partial y}+z\,\dfrac{\partial}{\partial z}.
\end{equation}
where $U(t)=a_0\,\exp\Big[-\dfrac{1}{2}\,\int\theta(t)\,dt\Big]$ and $W(t)=b_0\exp\Big[-\dfrac{1}{2}\,\int\theta(t)\,dt\Big]$ while $a_0$ and $b_0$ are arbitrary constants.

\textbf{Solution (GR4): ($\theta\,=\,0$):}
\begin{equation}  \label{u41-HS-12-GR}
\left\{
  \begin{array}{ll}
    X^0\,=\,a_1+\alpha\,t,\,\,\,\,\,\,\,\,\,\,\,\,\,\,\,\,\,\,\,\,\,\,\,\,\,\,\,\,\,\,\,\,\,\,\,
    X^1\,=\,a_3+a_2\,x,\\
    \\
X^2=a_4+a_5\,y+a_6\,z,
\,\,\,\,\,\,\,\,\,\,\,\,\,\,\,\,
X^3=a_7-a_6\,y+a_5\,z,
  \end{array}
\right.
\end{equation}
where
$U(t)=a_0\,\left(a_1+\alpha\,t\right)^{1-a_2/\alpha}$, $W(t)=b_0\,\left(a_1+\alpha\,t\right)^{1-a_5/\alpha}$ and $\theta(t)\,=\,0$ while $\alpha$, $a_i$, $i=0,1,...,7$ are arbitrary
constants such that $\alpha\,\neq\,0$. After removing a minor typing error in case (4) of \cite{shabbir9}, one can see that our obtained result is exactly same with their result.

\textbf{Solution (LG4): ($\theta\,=\,2\,\theta_0$):}
\begin{equation}  \label{u41-HS-12-C}
\left\{
  \begin{array}{ll}
    X^0\,=\,a_1\,e^{-\theta_0\,t}+\alpha\,\theta_0^{-1},\,\,\,\,\,\,\,\,\,\,\,\,\,\,\,\,
    X^1\,=\,a_3+a_2\,x,\\
    \\
X^2=a_4+a_5\,y+a_6\,z,
\,\,\,\,\,\,\,\,\,\,\,\,\,\,\,\,\,\,\,\,\,
X^3=a_7-a_6\,y+a_5\,z,
  \end{array}
\right.
\end{equation}
where $U(t)=a_0\,e^{-\theta_0\,t}\,\left(a_1+\alpha\,\theta_0^{-1}\,e^{\theta_0\,t}\right)^{1-a_2/\alpha}$ and

$W(t)=b_0\,e^{-\theta_0\,t}\,\left(a_1+\alpha\,\theta_0^{-1}\,e^{\theta_0\,t}\right)^{1-a_5/\alpha}$ while $a_0$ and $b_0$, $\alpha$, $a_i$, $i=1,...,7$ are arbitrary constants such that $\alpha\,\neq\,0$ while $\theta(t)$ is an arbitrary function.\\

\subsection{Killing Vector Fields}
In this sub-section, we are going to list all those cases where the spacetime under consideration
does not admit proper LHVFs and the HVFs are just the KVFs in the Lyra geometry.\\
\textbf{Solution (LG5):}
\begin{equation}  \label{u41-HS-1}
\left\{
  \begin{array}{ll}
    X^0\,=\,\dfrac{4\,(a_1+2\,a_2\,y+2\,a_3\,z)}{W_0(t)+2\,\theta(t)},\,\,\,\,\,\,\,\,\,\,\,\,\,\,\,\,\,\,\,\,
    X^1\,=\,a_4,\\
\\
X^2=a_5-a_1\,y+a_6\,z-2\,a_3\,y\,z-a_2\,\left(y^2-z^2-8\,\int\dfrac{W^{-2}(t)}{W_0(t)+2\,\theta(t)}\,dt\right),\\
\\
X^3=a_7-a_6\,y-a_1\,z-2\,a_2\,y\,z+a_3\,\left(y^2-z^2+8\,\int\dfrac{W^{-2}(t)}{W_0(t)+2\,\theta(t)}\,dt\right),
  \end{array}
\right.
\end{equation}
where $U(t)=a_0\,\exp\Big[-\dfrac{1}{2}\,\int\theta(t)\,dt\Big]$, $W(t)=b_0\,\exp\Big[\dfrac{1}{4}\,\int W_0(t)\,dt\Big]$,
$$W_0(t)=\Bigg(4\,b_1+\int\Big[\theta^2(t)-2\,\theta'(t)\Big]\,\exp\Big[-\dfrac{1}{2}\,\int\theta(t)\,dt\Big]\,dt\Bigg)\,\exp\Big[\dfrac{1}{2}\,\int\theta(t)\,dt\Big],$$
while $\theta(t)$ is an arbitrary function, $b_0$, $b_1$, $a_i$, $i=0,1,...,7$ are
arbitrary constants such that $\alpha\,=\,0$, $a_0\,\neq\,0$ and $b_0\,\neq\,0$. This result shows that for particular metric functions (as given above), Bianchi type I spacetime does not admit proper LHV.

\textbf{Solution (GR5): ($\theta\,=\,0$):}
\begin{equation}  \label{u41-HS-1-GR}
\left\{
  \begin{array}{ll}
    X^0\,=\,\dfrac{a_1+2\,a_2\,y+2\,a_3\,z}{b_1},\,\,\,\,\,\,\,\,\,\,\,\,\,\,\,\,\,\,\,\,
    X^1\,=\,a_4,\\
\\
X^2=a_5-a_1\,y+a_6\,z-2\,a_3\,y\,z-a_2\,\left(y^2-z^2+\dfrac{e^{-2\,b_1\,t}}{b_0^2\,b_1^2}\right),\\
\\
X^3=a_7-a_6\,y-a_1\,z-2\,a_2\,y\,z+a_3\,\left(y^2-z^2-\dfrac{e^{-2\,b_1\,t}}{b_0^2\,b_1^2}\right),
  \end{array}
\right.
\end{equation}
where $U(t)=a_0$ and $W(t)=b_0\,e^{b_1\,t}$ while $\theta(t)\,=\,0$, $b_0$, $b_1$, $a_i$, $i=0,1,...,7$ are
constants, while $\alpha\,=\,0$, $a_0\,\neq\,0$ and $b_0\,\neq\,0$.

\textbf{Solution (LG5): ($\theta\,=\,2\,\theta_0$):}
\begin{equation}  \label{u41-HS-1-GR}
\left\{
  \begin{array}{ll}
    X^0\,=\,\dfrac{e^{-\theta_0\,t}\,\left(a_1+2\,a_2\,y+2\,a_3\,z\right)}{b_1},\,\,\,\,\,\,\,\,\,\,\,\,\,\,\,\,\,\,\,\,
    X^1\,=\,a_4,\\
\\
X^2=a_5-a_1\,y+a_6\,z-2\,a_3\,y\,z-a_2\,\left(y^2-z^2+\dfrac{1}{b_0^2\,b_1^2}\,\exp\left[-\dfrac{2\,b_1}{\theta_0}\,e^{\theta_0\,t}\right]\right),\\
\\
X^3=a_7-a_6\,y-a_1\,z-2\,a_2\,y\,z+a_3\,\left(y^2-z^2-\dfrac{1}{b_0^2\,b_1^2}\,\exp\left[-\dfrac{2\,b_1}{\theta_0}\,e^{\theta_0\,t}\right]\right),
  \end{array}
\right.
\end{equation}
where $U(t)=a_0\,e^{-\theta_0\,t}$ and $W(t)=b_0\,e^{-\theta_0\,t}\,\exp\left[\dfrac{b_1}{\theta_0}\,e^{\theta_0\,t}\right]$ while $b_0$, $b_1$, $a_i$, $i=0,1,...,7$ are constants such that $\alpha\,=\,0$, $a_0\,\neq\,0$ and $b_0\,\neq\,0$.\\

\textbf{Solution (LG6):}
\begin{equation}  \label{u41-HS-5}
\left\{
  \begin{array}{ll}
    X^0\,=\,\dfrac{8\,(a_1+a_2\,x+a_3\,y+a_4\,z)}{W_0(t)+2\,\theta(t)},\\
\\
X^1\,=\,a_5-2\,x\,\left(a_1+a_3\,y+a_4\,z\right)+a_6\,y+a_7\,z\\
\,\,\,\,\,\,\,\,\,\,\,\,\,\,\,\,\,\,\,\,\,\,\,\,\,\,\,\,\,\,\,\,\,\,\,\,\,\,\,\,\,\,\,\,\,\,\,\,\,\,\,\,\,\,\,
-a_2\,\left[x^2-y^2-z^2-8\,\int\dfrac{W^{-2}(t)}{W_0(t)+2\,\theta(t)}\,dt\right],\\
\\
X^2\,=\,a_8-a_6\,x-2\,y\,\left(a_1+a_2\,x+a_4\,z\right)+a_9\,z\\
\,\,\,\,\,\,\,\,\,\,\,\,\,\,\,\,\,\,\,\,\,\,\,\,\,\,\,\,\,\,\,\,\,\,\,\,\,\,\,\,\,\,\,\,\,\,\,\,\,\,\,\,\,\,\,
+a_3\,\left[x^2-y^2+z^2+8\,\int\dfrac{W^{-2}(t)}{W_0(t)+2\,\theta(t)}\,dt\right],\\
\\
X^3\,=\,a_{10}-a_7\,x-a_9\,y-2\,z\,\left(a_1+a_2\,x+a_3\,y\right)\\
\,\,\,\,\,\,\,\,\,\,\,\,\,\,\,\,\,\,\,\,\,\,\,\,\,\,\,\,\,\,\,\,\,\,\,\,\,\,\,\,\,\,\,\,\,\,\,\,\,\,\,\,\,\,\,
+a_4\,\left[x^2+y^2-z^2+8\,\int\dfrac{W^{-2}(t)}{W_0(t)+2\,\theta(t)}\,dt\right],
  \end{array}
\right.
\end{equation}
where $U(t)\,=\,W(t)=\exp\Big[\dfrac{1}{4}\,\int W_0(t)\,dt\Big]$,
$$W_0(t)=\Bigg(4\,b_1+\int\Big[\theta^2(t)-2\,\theta'(t)\Big]\,\exp\Big[-\dfrac{1}{2}\,\int\theta(t)\,dt\Big]\,dt\Bigg)\,\exp\Big[\dfrac{1}{2}\,\int\theta(t)\,dt\Big],$$
while $\theta(t)$ is an arbitrary function, $b_1$, $a_i$,
$i=0,1,...,10$ are arbitrary constants such that $\alpha\,=\,0$.

\textbf{Solution (GR6): ($\theta\,=\,0$):}
\begin{equation}  \label{u41-HS-5-GR}
\left\{
  \begin{array}{ll}
    X^0\,=\,\dfrac{2\,(a_1+a_2\,x+a_3\,y+a_4\,z)}{b_1},\\
\\
X^1\,=\,a_5-2\,x\,\left(a_1+a_3\,y+a_4\,z\right)+a_6\,y+a_7\,z-a_2\,\left[x^2-y^2-z^2+\dfrac{e^{-2\,b_1\,t}}{b_1^2}\right],\\
\\
X^2\,=\,a_8-a_6\,x-2\,y\,\left(a_1+a_2\,x+a_4\,z\right)+a_9\,z+a_3\,\left[x^2-y^2+z^2-\dfrac{e^{-2\,b_1\,t}}{b_1^2}\right],\\
\\
X^3\,=\,a_{10}-a_7\,x-a_9\,y-2\,z\,\left(a_1+a_2\,x+a_3\,y\right)+a_4\,\left[x^2+y^2-z^2-\dfrac{e^{-2\,b_1\,t}}{b_1^2}\right],
  \end{array}
\right.
\end{equation}
where $U(t)=W(t)=e^{b_1\,t}$ while $\theta(t)\,=\,0$, $b_1$, $a_i$, $i=0,1,...,10$ are
arbitrary constants such that $\alpha\,=\,0$.

\textbf{Solution (LG6): ($\theta\,=\,2\,\theta_0$):}
\begin{equation}  \label{u41-HS-5-GR}
\left\{
  \begin{array}{ll}
    X^0\,=\,\dfrac{2\,e^{-\theta_0\,t}\,(a_1+a_2\,x+a_3\,y+a_4\,z)}{b_1},\\
\\
X^1\,=\,a_5-2\,x\,\left(a_1+a_3\,y+a_4\,z\right)+a_6\,y+a_7\,z-a_2\,\left(x^2-y^2-z^2+\dfrac{1}{b_1^2}\,\exp\left[-\dfrac{2\,b_1}{\theta_0}\,e^{-2\,b_1\,t}\right]\right),\\
\\
X^2\,=\,a_8-a_6\,x-2\,y\,\left(a_1+a_2\,x+a_4\,z\right)+a_9\,z+a_3\,\left(x^2-y^2+z^2-\dfrac{1}{b_1^2}\,\exp\left[-\dfrac{2\,b_1}{\theta_0}\,e^{-2\,b_1\,t}\right]\right),\\
\\
X^3\,=\,a_{10}-a_7\,x-a_9\,y-2\,z\,\left(a_1+a_2\,x+a_3\,y\right)+a_4\,\left(x^2+y^2-z^2-\dfrac{1}{b_1^2}\,\exp\left[-\dfrac{2\,b_1}{\theta_0}\,e^{-2\,b_1\,t}\right]\right),
  \end{array}
\right.
\end{equation}
where $U(t)\,=\,W(t)\,=\,e^{-\theta_0\,t}\,\exp\left[\dfrac{b_1}{\theta_0}\,e^{-2\,b_1\,t}\right]$ while $b_1$, $a_i$, $i=0,1,...,10$ are
arbitrary constants such that $\alpha\,=\,0$.\\

\textbf{Solution (LG7):}
\begin{equation}  \label{u41-HS-9}
\left\{
  \begin{array}{ll}
    X^0\,=\,\left[a_1\,f_1(x)+a_2\,f_2(x)\right]\,\exp\Big[-\int\dfrac{\theta(t)}{2}\,dt\Big],\\
\\
X^1\,=\,a_3+\left[a_2\,f_1(x)+\gamma_0^{-1}\,a_1\,f'_1(x)\right]\,f_3(t),\\
\\
X^2=a_4+a_5\,z,\,\,\,\,\,\,\,\,\,\,\,\,\,\,\,\,\,\,\,\,\,\,\,\,\,\,\,\,\,\,\,\,\,\,X^3=a_6-a_5\,y,
  \end{array}
\right.
\end{equation}
where $\alpha=0$, $U(t)=\sqrt{\dfrac{\gamma_0}{f'_3(t)}}\,\exp\Big[-\int\dfrac{\theta(t)}{2}\,dt\Big]$, $W(t)=b_0\,\exp\Big[-\int\dfrac{\theta(t)}{2}\,dt\Big]$, $\theta(t)$ is an arbitrary function and $b_0$, $\gamma_0$, $a_i$, $i=1,...,6$ are arbitrary constants such that
$\gamma_0\,\neq\,0$ and $b_0\,\neq\,0$. The functions $f_1(x)$, $f_2(x)$ and $f_3(t)$ lead to two solutions in the following cases:\\

\textbf{Case (1):} $f_1(x)\,=\,\cos\left[\gamma_0\,x\right]$, $f_2(x)\,=\,\sin\left[\gamma_0\,x\right]$ and $f_3(t)\,=\,\tanh\left[c_0+\gamma_0\,\int\exp\Big[\int\dfrac{\theta(t)}{2}\,dt\Big]\,dt\right]$, where $c_0$ is an arbitrary constant.\\

\textbf{Case (2):} $f_1(x)\,=\,\cosh\left[\gamma_0\,x\right]$, $f_2(x)\,=\,\sinh\left[\gamma_0\,x\right]$ and $f_3(t)\,=\,\tan\left[c_0+\gamma_0\,\int\exp\Big[\int\dfrac{\theta(t)}{2}\,dt\Big]\,dt\right]$, where $c_0$ is an arbitrary constant.\\

\textbf{Solution (GR7): ($\theta\,=\,0$):} The HVs of general relativity takes the form (\ref{u41-HS-9}) with $\theta\,=\,0$ such that:
$f_3(t)\,=\,\tanh\left[c_0+\gamma_0\,t\right]$ in case (1) and $f_3(t)\,=\,\tan\left[c_0+\gamma_0\,t\right]$ in case (2).\\

The line elements in two cases become:
\begin{equation}  \label{1u31-01}
ds_1^2\,=\,-dt^2+\cosh^2\left[c_0+\gamma_0\,t\right]\,dx^2+b_0^2\,\left(dy^2+dz^2\right),
\end{equation}
and
\begin{equation}  \label{1u31-01}
ds_2^2\,=\,-dt^2+\cos^2\left[c_0+\gamma_0\,t\right]\,dx^2+b_0^2\,\left(dy^2+dz^2\right).
\end{equation}
Also, the non-zero components of the energy momentum tensor $\left(T_{ij}\,=\,R_{ij}-\dfrac{R}{2}\,g_{ij}\right)$ are $T_{33}\,=\,T_{33}\,=\,\pm\,b_0^2\,\gamma_0^2$.

\textbf{Solution (LG7): ($\theta\,=\,2\,\theta_0$):}
\begin{equation}  \label{u41-HS-9-C}
\left\{
  \begin{array}{ll}
    X^0\,=\,\left[a_1\,f_1(x)+a_2\,f_2(x)\right]\,e^{-\theta_0\,t},\\
\\
X^1\,=\,a_3+\left[a_2\,f_1(x)+\gamma_0^{-1}\,a_1\,f'_1(x)\right]\,f_3(t),\\
\\
X^2=a_4+a_5\,z,\,\,\,\,\,\,\,\,\,\,\,\,\,\,\,\,\,\,\,\,\,\,\,\,\,\,\,\,\,\,\,\,\,\,X^3=a_6-a_5\,y,
  \end{array}
\right.
\end{equation}
where $W(t)\,=\,b_0\,\,e^{-\theta_0\,t}$ such that: $f_3(t)\,=\,\tanh\left[c_0+\dfrac{\gamma_0\,e^{\theta_0\,t}}{\theta_0}\right]$ and $U(t)=\cosh\left[c_0+\dfrac{\gamma_0\,e^{\theta_0\,t}}{\theta_0}\right]$ in case (1) and $f_3(t)\,=\,\tan\left[c_0+\dfrac{\gamma_0\,e^{\theta_0\,t}}{\theta_0}\right]$ and $U(t)=\cos\left[c_0+\dfrac{\gamma_0\,e^{\theta_0\,t}}{\theta_0}\right]$ in case (2).\\

\textbf{Solution (LG8):}
\begin{equation}  \label{u41-HS-11}
\left\{
  \begin{array}{ll}
X^0\,=\,\left(a_1-2\,b_0\,a_2\,x\right)\,W(t),\\
\\
X^1\,=\,a_3-b_0\,a_1\,x+a_2\,\left(b_0^2\,x^2+\exp\left[-2\,b_0\,\int \exp\left[\int\dfrac{\theta(t)}{2}\,dt\right]\,dt\right]\right),\\
\\
X^2\,=\,a_4+a_5\,z,\,\,\,\,\,\,\,\,\,\,\,\,\,\,\,\,\,\,\,\,\,\,\,\,\,\,\,\,
X^3=a_6-a_5\,y,
  \end{array}
\right.
\end{equation}
where $\alpha=0$,
$W(t)\,=\,\exp\left[-\int\dfrac{\theta(t)}{2}\,dt\right]$ and

$U(t)\,=\,W(t)\,\exp\left[b_0\,\int e^{\int\dfrac{\theta(t)}{2}\,dt}\,dt\right]$ while $\theta(t)$ is an arbitrary
function and $a_i$, $i=1,...,6$ are arbitrary constants such that $b_0\,\neq\,0$.\\

\textbf{Solution (GR8): ($\theta\,=\,0$):}
\begin{equation}  \label{u41-HS-11-GR}
\left\{
  \begin{array}{ll}
X^0\,=\,a_1-2\,b_0\,c_0\,a_2\,x,\\
\\
X^1\,=\,a_3-b_0\,a_1\,x+a_2\,\left(b_0^2\,x^2+e^{-2\,b_0\,t}\right),\\
\\
X^2\,=\,a_4+a_5\,z,\,\,\,\,\,\,\,\,\,\,\,\,\,\,\,\,\,\,\,\,\,\,\,\,\,\,\,\,
X^3=a_6-a_5\,y,
  \end{array}
\right.
\end{equation}
where $\alpha=0$,
$W(t)=1$ and $U(t)\,=\,e^{b_0\,t}$ while $a_i$, $i=1,...,6$ are arbitrary constants such that $b_0\,\neq\,0$.\\

\textbf{Solution (LG8): ($\theta\,=\,2\,\theta_0$):}
\begin{equation}  \label{u41-HS-11-C}
\left\{
  \begin{array}{ll}
X^0\,=\,\left(a_1-2\,b_0\,c_0\,a_2\,x\right)\,e^{-\theta_0\,t},\\
\\
X^1\,=\,a_3-b_0\,a_1\,x+a_2\,\left(b_0^2\,x^2+\exp\left[-\dfrac{2\,b_0}{\theta_0}\,e^{\theta_0\,t}\right]\right),\\
\\
X^2\,=\,a_4+a_5\,z,\,\,\,\,\,\,\,\,\,\,\,\,\,\,\,\,\,\,\,\,\,\,\,\,\,\,\,\,
X^3=a_6-a_5\,y,
  \end{array}
\right.
\end{equation}
where $\alpha=0$,
$W(t)=e^{-\theta_0\,t}$ and $U(t)\,=\,e^{-\theta_0\,t}\,\exp\left[\dfrac{b_0}{\theta_0}\,e^{\theta_0\,t}\right]$ while $a_i$, $i=1,...,6$ are arbitrary constants such that $b_0\,\neq\,0$.\\

\section{Matter Collineation}

A vector field is said to be a Matter Collineations (MC), if the Lie derivative of energy-momentum tensor vanishes along it, mathematically
\begin{equation}\label{u11-MC-1}
\mathcal{L}_{X}\,T_{ij}\,=\,0.
\end{equation}

If $X$ is a HV then also $\mathcal{L}_{X}\,T_{ij}\,=\,0$. Thus every HV is a MC also but converse does not hold in general. The authors of the paper \cite{gad3}, claimed that in Lyra geometry, barotropic equation of state
\begin{equation}\label{u11-MC-20}
\rho(t)\,=\gamma\,p(t),
\end{equation}

is never satisfied when $\theta$ is taken as function of $t$ or a constant. They claimed that Eq. of the form (\ref{u11-MC-20}) satisfies only when $\theta=0$. In order to check whether or not Eq. (\ref{u11-MC-20}) satisfies in our cases, we shall try to establish a relation between the density and pressure. For this purpose we take the matter field for spacetime under consideration as a perfect fluid, that is, taking energy-momentum tensor of the form
\begin{equation}\label{u11-MC-2}
T_{ij}\,=\,(\rho+p)\,u_{i}\,u_{j}-p\,g_{ij},
\end{equation}
where, for the spacetime (\ref{u31}), the four-velocity vector is taken as $u^{i}=(1,0,0,0)$, $u^{i}\,u_{i}\,=\,-1$. The non-zero components of the energy-momentum tensor (\ref{u11-MC-2}) is given by;
\begin{equation}\label{u11-MC-3}
T_{00}\,=\,\rho(t),\,\,\,\,\,\,\,\,\,\,T_{11}\,=\,U^2(t)\,p(t),\,\,\,\,\,\,\,\,\,T_{22}\,=\,T_{33}\,=\,W^2(t)\,p(t).
\end{equation}

Substituting the proper Lyra homothetic vector (\ref{u41-HS-13-P}) of the solution \textbf{(LG1)} in the MC equation (\ref{u11-MC-1}) and making use of (\ref{u11-MC-3}), we obtain the following constrains:
\begin{equation}  \label{u11-MC-LG1}
\left\{
  \begin{array}{ll}
    a_2\,=\,a_3\,=\,a_4\,=\,0,\\
    \\
    2\,\alpha\,\rho(t)\,\exp\left[\int\dfrac{\theta(t)}{2}\,dt\right]\,=\,
    \left(a_1+\alpha\int\,\exp\left[\int\dfrac{\theta(t)}{2}\,dt\right]\,dt\right)\,\left[\rho(t)\,\theta(t)-\rho'(t)\right],\\
    \\
   2\,\alpha\,p(t)\,\exp\left[\int\dfrac{\theta(t)}{2}\,dt\right]\,=\,
    \left(a_1+\alpha\int\,\exp\left[\int\dfrac{\theta(t)}{2}\,dt\right]\,dt\right)\,\left[p(t)\,\theta(t)-p'(t)\right].
  \end{array}
\right.
\end{equation}
Integrating the above equations with respect to $t$, we get:
\begin{equation}  \label{u11-MC-LG1-01}
\left\{
  \begin{array}{ll}
    \rho(t)\,=\,\rho_0\,e^{\int\theta(t)\,dt}\,\left(a_1+\alpha\int\,\exp\left[\int\dfrac{\theta(t)}{2}\,dt\right]\,dt\right)^{-2},\\
    \\
    p(t)\,=\,p_0\,e^{\int\theta(t)\,dt}\,\left(a_1+\alpha\int\,\exp\left[\int\dfrac{\theta(t)}{2}\,dt\right]\,dt\right)^{-2},
  \end{array}
\right.
\end{equation}
where $\rho_0$ and $p_0$ are constants of integration.

Looking at the values of $p(t)$ and $\rho(t)$ it is observed that when $\theta=\theta(t)$, the barotropic equation of state $\rho(t)\,=\,\gamma\,p(t)$,
($\gamma=$ constant, $0\,\leq\,\gamma\,\leq\,1$) is satisfied for $\gamma\,=\,\rho_0/p_0$. It is interesting to see when $\theta(t)\,=\,0$ and $\theta(t)\,=\,2\,\theta_0$, the pressure and density are given respectively by:
\begin{equation}  \label{u11-MC-LG1-02}
\left\{
  \begin{array}{ll}
    \theta(t)\,=\,0\,\,\Rightarrow\,\,\rho(t)\,=\,\rho_0\,\left(a_1+\alpha\,t\right)^{-2},\,\,\,\,\,\,\,p(t)\,=\,p_0\,\left(a_1+\alpha\,t\right)^{-2},\\
    \\
    \theta(t)\,=\,2\,\theta_0\,\,\Rightarrow\,\,\rho(t)\,=\,\rho_0\,\theta_0^2\,\left(\alpha+a_1\,\theta_0\,e^{-\theta_0\,t}\right)^{-2},\,\,\,\,\,\,\,
    p(t)\,=\,p_0\,\rho(t)/\rho_0.
  \end{array}
\right.
\end{equation}
It is important to observe that when $\theta=0$, both pressure and density of the matter field are functions of $t$ and also depends upon the homothetic factor $\alpha$. If we put the homothetic factor equal to zero then both the pressure and density will become constant. Also, for all the remaining proper Lyra homothetic vectors similar results were obtained, which we are going to omit here.

Next we put the Lyra Killing vector fields of the solution \textbf{(LG5)} in the MC equation (\ref{u11-MC-1}), we find the following constrains:
\begin{equation}  \label{u11-MC-LG4}
\left\{
  \begin{array}{ll}
    a_2\,=\,a_3\,=\,0,\,\,\,\,\,\,\,
    a_1\,\left[\rho'(t)-\theta(t)\,\rho(t)\right]\,=\,0,\,\,\,\,\,\,\,
    a_1\,\left[p'(t)-\theta(t)\,p(t)\right]\,=\,0.
  \end{array}
\right.
\end{equation}
The above equation admits two solutions as the following:

\textbf{(1):} $\,a_1\,=\,0$, $\rho(t)$ and $p(t)$ are both functions of $t$.

\textbf{(2):} $a_1\,\neq\,0$, $\rho(t)\,=\,\rho_0\,e^{\int\,\theta(t)\,dt}$ and $p(t)\,=\,p_0\,e^{\int\,\theta(t)\,dt}$, where $p_0$ and $\rho_0$ are arbitrary constants.

Looking at the values of $p(t)$ and $\rho(t)$ it is observed that when $\theta=\theta(t)$, the barotropic equation of state $\rho(t)\,=\,\gamma\,p(t)$,
($\gamma=$ constant, $0\,\leq\,\gamma\,\leq\,1$) is satisfied for $\gamma\,=\,\rho_0/p_0$. It is interesting to see when $\theta(t)\,=\,0$ and $\theta(t)\,=\,2\,\theta_0$, the metric functions, the pressure and density are given, respectively by:
\begin{equation}  \label{u11-MC-LG1-03}
\left\{
  \begin{array}{ll}
    \theta(t)\,=\,0\,\,\Rightarrow\,\,U(t)\,=\,a_0,\,\,\,W(t)\,=\,b_0\,e^{-b_1\,t},\,\,\,\rho(t)\,=\,\rho_0,\,\,\,p(t)\,=\,p_0,\\
    \\
    \theta(t)\,=\,2\,\theta_0\,\,\Rightarrow\,\,U(t)\,=\,a_0\,e^{-\theta_0\,t},\,\,\,W(t)\,=\,b_0\,e^{-\theta_0\,t}\,\exp\left[\dfrac{b_1}{\theta_0}\,e^{\theta_0\,t}\right],\\
    \rho(t)\,=\,\rho_0\,e^{2\,\theta_0\,t},\,\,\,\,\,\,\,\,\,p(t)\,=\,p_0\,e^{2\,\theta_0\,t}.
  \end{array}
\right.
\end{equation}
In all the remaining cases of Lyra Killing vector fields similar results were obtained, which we are going to omit here.

\section{Conclusion}

In this paper Bianchi type I spacetimes are classified according to
their homothetic vectors in the context of Lyra geometry.
Our classification reveals that there exist only three different possibilities
where the spacetime under consideration admit proper LHVs for the special choice of
the metric functions. Also the cases where Bianchi type I spacetimes do not admit proper
LHVs and LHVs are just LKVs are explored. In all these cases the metric functions
are obtained and it comes out that these metric functions are dependent upon
the displacement vector $\theta(t)$. For the most general case \textbf{(LG1)}, we have given the Lie algebra
structure, which is closed. In order to obtain homothetic vectors of Bianchi type I spacetimes in
the context of general relativity, the displacement vector is set equal to zero in the
HVF components of Lyra geometry and in the metric functions.
In this way a classification of Bianchi type I spacetime according to homothetic vectors in general relativity
is also obtained.\\
In some recent papers \cite{gad2, gad3} it was claimed that barotropic equation of state (\ref{u11-MC-20})
never satisfies in context of Lyra geometry for plane symmetric and Bianchi type I spacetimes when the
displacement vector is a function of time or when it is a constant. In order to check the consistency
of those judgments, we took the matter field as a perfect fluid and explained the matter collineation symmetry for the spacetime under consideration.
We found that every LHV is also a matter collineation vector which in tern created the possibility to obtain a barotropic equation
of state of the form (\ref{u11-MC-20}).\\
Contrary to \cite{gad2, gad3}, in the case of Bianchi type-I spacetimes, we found that a barotropic equation of state of the form  (\ref{u11-MC-20}) is always possible
to form, when the displacement vector is a function of t or is a constant.\\



\begin{thebibliography}{99}

\bibitem{ali1} Ali AT, Yadav AK and Mahmood S,  \textit{Astr Space Sci} \textbf{349(1)},
539, (2014).

\bibitem{ali2} Ali AT, Yadav AK, Rahaman F and Mallick A,  \textit{Phys Scr} \textbf{89(11)},
115206, (2014).

\bibitem{att1} Attallah SK, El-Sabbagh MF and Ali AT, \textit{Commun Nonlinear Sci Numer Simulat} \textbf{12(7)}, 1153, (2007).

\bibitem{cole1} Coley AA and Tupper BOJ, \textit{Class Quantum Gravity} \textbf{11}, 2553, (1994).

\bibitem{gad1} Gad RM and Hassan MM,  \textit{Il Nuovo Cimento B} \textbf{12},
759, (2003).

\bibitem{gad2} Gad RM and Alofi AS,  \textit{Mod Phys Lett A} \textbf{22},
1450116, (2014).

\bibitem{gad3} Gad RM,  \textit{Int J Theor Phys} DOI 10.1007/s10773-015-2528-z, (2015).

\bibitem{hall1} Hall GS: "Symmetries and Curvature Structure in General Relativity", World Scientific, Singapore (2004).

\bibitem{katzin1} Katzin GH, Levine J and Davis WR, \textit{J Math Phys} \textbf{10}, 617, (1969).

\bibitem{katzin2} Katzin GH, Levine J and Davis WR, \textit{J Math Phys} \textbf{11}, 1518, (1970).

\bibitem{khan0} Khan H, Qadir A, Saifullah K and Ziad M, \textit{Eur Phys J Plus} \textbf{128}, 144, (2013).

\bibitem{khan1} Khan S, Hussain T and Khan GA, \textit{Rom J Phys} \textbf{59}, 488, (2014).

\bibitem{khan2} Khan S, Hussain T and Khan GA, \textit{Eur Phys J Plus} \textbf{129}, 228, (2014).

\bibitem{khan3} Khan S, Hussain T, Bokhari AH , Khan GA, \textit{Eur Phys J C} \textbf{75}, 523, (2015).

\bibitem{lyra1} Lyra G, \textit{Math Z} \textbf{54}, 52, (1951).

\bibitem{maar1} Maartens R, Maharaj SD and Tupper BOJ, \textit{J Math Phys} \textbf{27}, 2987, (1986).

\bibitem{olve1} Olver PJ: "Application of Lie Groups to
differential equations in graduate texts in mathematics", Vol. 107,
Second edition, Springer, New York, (1993).

\bibitem{ovsi1} Ovsiannikov LV: "Group analysis of
differential equations", Translated by Chapovsky, Y. Ed. Ames W.F.,
Academic Press, New York-London (1982).

\bibitem{sen1} Sen DK, \textit{Zeitschrift f\"{u}r Physik
C: Particles and Fields} \textbf{149}, 311, (1957).

\bibitem{sen2} Sen DK, \textit{Canad Math Soc} \textbf{3}, 255, (1960).

\bibitem{sen3} Sen DK and Dunn KA, \textit{J Math Phys} \textbf{12}, 578, (1971).

\bibitem{shabbir1} Shabbir G, Khan S and Ali A, \textit{Commun Theor Phys} \textbf{55}, 268, (2011).

\bibitem{shabbir2} Shabbir G, Ali A and Khan S, \textit{Chin Phys B} \textbf{20}, 070401, (2011).

\bibitem{shabbir3} Shabbir G, Khan S and Amir MJ, \textit{Braz J Phys} \textbf{41}, 184, (2011).

\bibitem{shabbir4} Shabbir G and Khan S, \textit{Rom J Phys} \textbf{57}, 571, (2012).

\bibitem{shabbir5} Shabbir G and Khan S, \textit{Commun Theor Phys} \textbf{54}, 675, (2010).

\bibitem{shabbir6} Shabbir G and Khan S, \textit{Mod Phys Lett A} \textbf{52}, 2145, (2010).

\bibitem{shabbir7} Shabbir G, Khan A and Khan S, \textit{Int J Theor Phys} \textbf{52}, 1182, (2013).

\bibitem{shabbir8} Shabbir G and Khan H, \textit{Rom J Phys} \textbf{59}, 79, (2014).

\bibitem{sharif1} Sharif M and Amir MJ, \textit{Mod Phys Lett} \textbf{23}, 963, (2008).

\bibitem{sharif2} Sharif M and Majeed B, \textit{Commun Theor Phys} \textbf{52}, 435, (2009).

\bibitem{steph1} Stephani H, Kramer D, MacCullam MAH, Hoenselears C and Herlt E: "Exact Solutions of Einstein's Field Equations", Cambridge University Press, Cambridge, (2003).

\bibitem{tsam1} Tsamparlis M, Paliathanasis A and Karpathopoulos L, \textit{Gen Relativ Gravit} \textbf{47}, 15, (2015).

\bibitem{tsam2} Tsamparlis M, \textit{Class Quantum Gravity} \textbf{15}, 2901, (1998).

\bibitem{weit1} Weitzenb\"{o}ck R: "Invariant en Theorie", Noordhoff, Groningen, (1923).

\bibitem{yadav1} Yadav AK and Ali AT,  \textit{Eur Phys J Plus} \textbf{129},
179, (2014).
\bibitem{shabbir9} Shabbir G and Amur KB, \textit{Appl Sci} \textbf{08}, 153, (2006).
\bibitem{ali3} Ali AT, arXiv:1512.04427v1 [physics.gen-ph]

\end{thebibliography}
\end{document}